\documentclass[pra,twocolumn,showpacs,showkeys]{revtex4}
\usepackage{graphics}
\usepackage{amsmath}
\usepackage{mathrsfs}
\usepackage{theorem}
\usepackage{overpic}
\usepackage{hyperref}
\usepackage{rotating}
\usepackage{amssymb}
\usepackage[english]{babel}
\usepackage{color,times}
\usepackage{CJK}

\newcommand{\ket}[1]{|#1\rangle}
\newcommand{\bra}[1]{\langle #1|}
\newcommand{\inp}[2]{\langle #1 | #2\rangle}
\newtheorem{theorem}{Conjecture}
\newtheorem{theorem1}{Definition}

\begin{document}
\title{A Generalized Geometric Measurement of Quantum Discord: Exact Treatment }
\author{Hai-Tao Cui}
\email{cuiht@aynu.edu.cn}
\author{Jun-Long Tian and Gui Yang}
\affiliation{School of Physics and Electric Engineering, Anyang Normal University, Anyang 455000, China}

\begin{abstract}
A generalization of the geometric measure of quantum discord is introduced in this article, based on Hellinger distance. Our definition has virtues of computability and independence of local measurement. In addition it also does not suffer from the recently raised critiques about quantum discord. The exact result can be obtained for bipartite pure states with arbitrary levels, which is completely determined by  the Schmidt decomposition. For bipartite mixed states the exact result can also be found for a special case. Furthermore the generalization into multipartite case is direct. It is shown that it can be evaluated exactly when the measured state is  invariant under permutation or translation. In addition the detection of quantum phase transition is also discussed for Lipkin-Meshkov-Glick and Dicke model.
\end{abstract}
\maketitle

\section{introduction}
Quantum systems can exhibit non-classical correlation by a different way from quantum entanglement, which is well-known quantum discord (QD). QD characterizes the minimal perturbation induced by single-party von Neumann measurement \cite{qd}. Thus there exists non-entangled state with non-zero QD. Recently QD has been shown as a resource to speed up the quantum information processing. For instance the determined quantum computation with one qubit  \cite{dqc1} and quantum metrology with noised states \cite{qm} have been demonstrated the advantage over classical computation, in which the quantum entanglement is not involved. In addition there exists extensive interest for QD in other diverse contexts \cite{review}.

However QD is  difficult to determine analytically because of the optimization in the definition. There exist very few exact results, even for the  two-qubit cases \cite{luo}. ( Recently an exact evaluation of QD is proposed in Ref \cite{ali10}. However it is pointed out in \cite{huang13} that this approach is not completely correct.) In addition  the computation of quantum discord is shown to be NP-complete \cite{huang14}; the running time of the computation of QD is increased exponentially with the dimension of  Hilbert space. Thus one has to find an efficient way to calculate QD.  Recently  geometric discord (GD) is introduced by Daki\'{c} and the coauthors, which is defined as the square form of the shortest distance between the measured state $\rho$ and zero-discord state $\chi$ in Hilbert space \cite{dakic}. Hence the optimization in the definition of geometric discord (GD) can be reduced greatly by the geometry of  $\rho$. Moreover a tight lower bound of GD can also been obtained for arbitrary states \cite{luo10}. Unfortunately  the square form of GD  is not monotonic under local operations; the value of GD can increase by local operations \cite{dakic, pz}. This deficit raises the question whether GD or QD could unambiguously manifest the non-classical correlation in quantum systmes \cite{pz, gheorghiu14}. In order to solve this problem, many generalizations of GD have been proposed. For instance the  rescaled GD is defined by rescaling the density operator with its norm  \cite{tufarelli}. Furthermore the so-called Schatten $p$-norm is also introduced to qualify the distance \cite{pnorm}, instead of the $2$-norm form used in Ref. \cite{dakic}. In addition the Bures distance is also introduced \cite{bures}. However  it is still difficult generally to find the analytical expression for GD since the  comple evaluation or optimization. Recently the Hellinger distance is introduced to measure QD \cite{chang13, luo04, dajka11}. This definition has a simple structure and can be evaluated readily. Moreover it is  monotonically nonincreasing (contractivity) under local operations \cite{luo04, dajka11}.

It is an interesting issue how to generalize QD into multipartite case. A direct way is to introduce the three-tangle of QD in tripartite case \cite{3tangle} and four-qubit case \cite{bai13}.  However it still is  difficult to determine analytically and furthermore to generalize into the case with more parties. Another alternative way is to find the minimum QD between any single party and the others \cite{ggd}, named as the global QD. However it does not include the other possible bipartite correlation and thus is not a comprehensive measurement of QD. Moreover the exact treatment of global QD is also difficult since one has to find optimal single-party von Neumann measurements for every party. A geometric generalization of global QD is introduced by finding the shortest distance from the zero global QD state \cite{xu12}. However the author adopt the $2$-norm of distance, which suffers from the problem of non-contractivity under local operations \cite{pz}.

With respect of these facts, we present an alternative approaching to QD in this article  by a generalization of Hellinger distance \cite{chang13, luo04, dajka11}.  The main idea is to  find the shortest distance from the completely classical state \cite{ll08}. Hence this approach is independent on the local measurements. It should be pointed out that our way is a multipartite generalization of the method in Ref. \cite{chang13},  with consideration of the recently arising critiques about QD \cite{pz,gheorghiu14}. By this way the exact evaluation can be reached for any bipartite pure state and also for some special mixed states. As for multipartite case the exact results can also be founded for symmetric states. This article is divided into several sections. The definition is presented in Section II, and a general expression  is also presented. Then Section III presents the exact evaluation for bipartite pure states and for a  special type of mixed states. In Section IV the situation for multipartite states is discussed and the exact result can be obtained for symmetric state. In addition we also show its ability of marking the quantum phase transition in many-body systems in Section V. In Section VI a discussion of multilevel case is presented in comparison with the studies in Ref.\cite{chang13}. Conclusion and discussion are presented in final section.

\section{definition and technical preparation}
We first present the definition.
\begin{theorem1}\label{def}
Given arbitrary state $\rho$ and completely classical state $\sigma$, we define geometric measure of QD as
\begin{eqnarray}\label{gqd}
D^{H}=\frac{1}{2}\min_{\sigma}\parallel \sqrt{\rho}-\sqrt{\sigma}\parallel^2,
\end{eqnarray}
where $\parallel \cdot \parallel$ is the Hilbert-Schmidt norm and the superscript means the Hellinger distance.
\end{theorem1}
$\sigma$ can be written as the  probabilistic mixture of local distinguishable states
\begin{eqnarray}
\sigma=\sum p_{k_1, k_2, \cdots, k_n}\ket{k_1, k_2, \cdots, k_n}\bra{k_1, k_2, \cdots, k_n},
\end{eqnarray}
in which $p_{k_1, k_2, \cdots, k_n}$ is a joint probability distribution and local states $\ket{k_i}$ span an orthonormal basis \cite{modi2010}. The correlations in $\sigma$ are identified as classical \cite{qd, ll08}.  For qubit case, a general orthonormal basis can be constructed by   $\ket{+}_i=\cos \frac{\theta_i}{2} \ket{1}+\exp(-i\phi_i)\sin \frac{\theta_i}{2}\ket{0}$ and $\ket{-}=\exp(i\phi_i)\sin \frac{\theta_i}{2}\ket{1}-\cos \frac{\theta_i}{2}\ket{0}$ with $\theta_i\in[0, \pi]$ and $\phi_i\in[0, 2\pi)$.

By Eq. \eqref{gqd},
\begin{eqnarray} \label{drho}
D^{H}=1- \max_{\sigma}\text{Tr} \left[\sqrt{\rho}\sqrt{\sigma}\right].
\end{eqnarray}
Then the evaluation of $D^{H}(\rho)$ is reduced to find the maximal overlap of $\sqrt{\rho}$ and $\sqrt{\sigma}$.  For clarity $\sigma$ is rewritten as
\begin{eqnarray}
\sigma=\sum_n p_n \ket{\sigma_n}\bra{\sigma_n},
\end{eqnarray}
in which $\{\ket{\sigma_n}\}$ denotes the local basis $\{\ket{k_1, k_2, \cdots, k_n}\}$ and the probability  $p_n$ is to be determined. In the following  we first present the  general expressions for pure and mixed state cases respectively. The detailed discussion is presented in the next section. For convenience we set $d$ as the dimension of the local basis and  $p_d=1-\sum_{n=1}^{d-1}p_n$ with normalization.

-\emph{Pure state}- For a pure state  $\ket{\psi}$ , one gets
\begin{eqnarray}
D^{H}=1- \max_{\{p_n, \sigma_n\}} \sum_n \left|\inp{\psi}{\sigma_n}\right|^2\sqrt{p_n}.
\end{eqnarray}
With respect of  $p_n$, the extremal values  of $D^H$ appear when
\begin{eqnarray}
&&\frac{\partial D^{H}}{\partial p_i}=0\Rightarrow\nonumber\\
&&\frac{\left|\inp{\psi}{\sigma_i}\right|^2}{\sqrt{p_i}}=\frac{\left|\inp{\psi}{\sigma_N}\right|^2}{\sqrt{p_N}},i=1,2,\cdots,d-1.
\end{eqnarray}
It is not difficult to find $\frac{\partial^2 D^{H}(\rho)}{\partial p_i\partial p_j}>0$.  Then the minimal extreme  satisfies the relation
\begin{eqnarray}\label{bpp}
p_i= \frac{\left|\inp{\psi}{\sigma_i}\right|^4}{\sum_n \left|\inp{\psi}{\sigma_n}\right|^4},  i=1,2,\cdots,d.
\end{eqnarray}
Consequently $D^{H}$ reduces to
\begin{eqnarray}\label{bp}
D^{H}=1 - \max_{\{\sigma_n\}} \sqrt{\sum_n \left|\inp{\psi}{\sigma_n}\right|^4}.
\end{eqnarray}

-\emph{Mixed state}- With the spectrum decomposition $\rho=\sum_k \lambda_k\ket{\phi_k}\bra{\phi_k}$, Eq.\eqref{drho} can be written as
\begin{eqnarray}
D^{H}=1 -   \max_{\{p_n, \sigma_n\}} \sum_{n, k}\sqrt{\lambda_k}\sqrt{p_n}\left|\inp{\phi_k}{\sigma_n}\right|^2.
\end{eqnarray}
With respect  of $p_i$, the extremal points can be decided by
\begin{eqnarray}
&&\frac{\partial D^{H}}{\partial p_i}=0\nonumber\\
&&\Rightarrow\frac{\sum_{k}\sqrt{\lambda_k}\left|\inp{\phi_k}{\sigma_i}\right|^2}{\sqrt{p_i}}
=\frac{\sum_{k}\sqrt{\lambda_k}\left|\inp{\phi_k}{\sigma_N}\right|^2}{\sqrt{p_N}},\nonumber\\
&& i=1,2,\cdots,N-1.
\end{eqnarray}
Directly $\frac{\partial^2 D^{H}}{\partial p_i\partial p_j}>0$.  Similarly one can obtain the relation
 \begin{eqnarray}\label{bmp}
p_i= \frac{\left(\sum_{k}\sqrt{\lambda_k}\left|\inp{\phi_k}{\sigma_i}\right|^2\right)^2}{\sum_n \left(\sum_{k}\sqrt{\lambda_k}\left|\inp{\phi_k}{\sigma_n}\right|^2\right)^2},  i=1,2,\cdots,d.
\end{eqnarray}
Then  $D^{H}$ reduces to
\begin{eqnarray}\label{bm}
D^{H}=1-\max_{\{\sigma_n\}} \sqrt{\sum_n \left(\sum_{k}\sqrt{\lambda_k}\left|\inp{\phi_k}{\sigma_n}\right|^2\right)^2},
\end{eqnarray}
The determination of $\sigma_n$  depends on $\rho$. It should be pointed out that the expressions above are general. However for simplicity the following studies  would focus mainly on qubit system since the simplicity and interest in quantum information processing. The extension into multi-level state will be presented in Sec. VI.

\section{bipartite state: exact treatment}

\emph{-Pure case-}  It is well known that  any bipartite pure state $\ket{\psi}_{AB}$ can be written in a concise form  $\ket{\psi}_{AB}=\sum_i \lambda_i \ket{i_A}\ket{i_B}$ by Schmidt decomposition.  The Schmidt bases $\ket{i_A}$ and $\ket{i_B}$ span a special space, of which the dimension is the minimal needed to expand $\ket{\psi}_{AB}$ by local orthonormal states. Hence in order to find the maximal overlap of $\ket{\psi}_{AB}$ and $\sigma$, it is a natural conjecture that $\sigma$ should belong to the special space too, i.e.,
\begin{eqnarray}
\sigma=\sum_i p_i\ket{i_A, i_B}\bra{i_A, i_B},
\end{eqnarray}
in which $p_i=\frac{|\lambda_i|^4}{\sum_n |\lambda_n|^4}$ by Eq.\eqref{bpp}.

However we do not know how to prove this conjecture exactly. An example is presented in order to display validity of the statement. We try to find $D^{H}$ for
\begin{eqnarray}
\ket{\psi}_{AB}&=&\frac{1}{\sqrt{2}}\ket{1}_A\left(\frac{1}{2}\ket{1}_B+\frac{\sqrt{3}}{2}\ket{0}_B\right)+\nonumber\\
&&\frac{1}{\sqrt{2}}\ket{0}_A\left(\frac{\sqrt{3}}{2}\ket{1}_B+\frac{1}{2}\ket{0}_B\right),
\end{eqnarray}
of which  Schmidt decomposition is
\begin{eqnarray}
\ket{\psi}_{AB}&=&\frac{\sqrt{2 + \sqrt{3}}}{2}\frac{\ket{1}_A+\ket{0}_A}{\sqrt{2}}\frac{\ket{1}_B+\ket{0}_B}{\sqrt{2}}\nonumber\\&+&
\frac{\sqrt{2 - \sqrt{3}}}{2}\frac{\ket{1}_A-\ket{0}_A}{\sqrt{2}}\frac{\ket{1}_B-\ket{0}_B}{\sqrt{2}}.
\end{eqnarray}
By Eq. \eqref{bp},  it is reduced to find the maximum of the overlap
\begin{eqnarray}
&&\sum_n \left|{_{AB}\inp{\psi}{\sigma_n}}\right|^4\nonumber\\
&=&(\tfrac{1}{4}-c_0+c_1+c_2)^2+ (\tfrac{1}{4}+c_0+c_1-c_2)^2\nonumber\\
&&+ (\tfrac{1}{4}+c_0-c_1+c_2)^2+(\tfrac{1}{4}-c_0-c_1-c_2)^2\nonumber\\
&=&4\left(\tfrac{1}{16}+c^2_0+c^2_1+c^2_2\right)
\end{eqnarray}
in which,
\begin{eqnarray}
c_0&=&\tfrac{1}{8}\left\{\cos\theta_1 \cos\theta_2 \right.\nonumber\\
&-&\left.\tfrac{1}{2}\sin\theta_1 \sin\theta_2 \left[\cos(\phi_1+\phi_2)
+3\cos(\phi_1-\phi_2)\right]\right\}\nonumber\\
c_1&=&\tfrac{\sqrt{3}}{8}\sin\theta_1 \cos\phi_1\nonumber\\
c_2&=&\tfrac{\sqrt{3}}{8}\sin\theta_2 \cos\phi_2
\end{eqnarray}
By analysis $c^2_0$ have the maximal value of 1/16 when $\theta_1=\theta_2=\pi/2$ and $\phi_{1(2)}=0, \pi$. Meanwhile $c^2_1$ and $c^2_2$ has the maximal values too.

It is obvious that the formula is invariant for $\theta_{1(2)}\leftrightarrow \pi- \theta_{1(2)}$. Consequently its extremal value happens  when $\theta_{1(2)}=0, \pi, \pi/2$. Then
\begin{eqnarray}
\sum_n \left|{_{AB}\inp{\psi}{\sigma_n}}\right|^4=\tfrac{7}{8}=\left(\tfrac{\sqrt{2+\sqrt{3}}}{2}\right)^4+\left(\tfrac{\sqrt{2-\sqrt{3}}}{2}\right)^4,
\end{eqnarray}
which obviously is the sum of the fourth power of the Schmidt coefficients. So $D^{H}=1-\sqrt{\tfrac{7}{8}}$. The "nearest" $\sigma$ is
\begin{eqnarray}
\sigma_{AB}&=&\tfrac{7+4\sqrt{3}}{14} \ket{1_x}_A\bra{1_x}\otimes\ket{1_x}_B\bra{1_x}\nonumber\\
&&+\tfrac{7-4\sqrt{3}}{14} \ket{0_x}_A\bra{0_x}\otimes\ket{0_x}_B\bra{0_x},
\end{eqnarray}
in which
$\ket{1_x}=\tfrac{1}{\sqrt{2}}(\ket{1}+\ket{0})$ and $\ket{0_x}=\tfrac{1}{\sqrt{2}}(\ket{1}-\ket{0})$.

By this example we obtain the first conjecture
\begin{theorem}\label{c1}
For arbitrary pure bipartite state, which has Schmidt decomposition
\begin{eqnarray*}
 \ket{\psi}_{AB}=\sum_i \lambda_i\ket{i_A} \ket{i_B},
\end{eqnarray*}
the "nearest" completely classical state $\sigma$ can be written as
\begin{eqnarray}
\sigma=\frac{1}{\lambda}\sum_{i}\left|\lambda_i\right|^4\ket{i_A}\bra{i_A}\otimes\ket{i_B}\bra{i_B},
\end{eqnarray}
in which $\lambda=\sum_i\left|\lambda_i\right|^4$. Then
\begin{eqnarray}
 D^{H}=1- \sqrt{\lambda}
\end{eqnarray}
\end{theorem}

\emph{-Mixed case-} As for spectrum decomposition $\rho=\sum_k \lambda_k\ket{\phi_k}\bra{\phi_k}$,  we cannot find a general result  for  $D^{H}$ since the eigenstates of $\ket{\phi_k}$ does not necessarily share the same Schmidt bases. However it is still possible for a exact treatment  when $\rho$ shows "X" form
\begin{eqnarray}\label{xstate}
\rho_X=\left(\begin{array}{ccccccc}
\rho_{11}&0&\cdots&\cdots&\cdots&0&\rho_{1n} \\
0&\rho_{22}&0&\cdots&0&\rho_{2(n-1)}&0\\
\vdots&  &\ddots& &{\begin{rotate}{90}$\ddots$\end{rotate}}&  &\vdots\\
\vdots&  &{\begin{rotate}{90}$\ddots$\end{rotate}}& & \ddots&  &\vdots\\
0&\rho_{(n-1)2}&0&\cdots&0&\rho_{(n-1)(n-1)}&0\\
\rho_{n1}&0&\cdots&\cdots&\cdots&0&\rho_{nn}
\end{array}\right),
\end{eqnarray}
in which the orthonormal basis is spanned by local states $\ket{i}_A\ket{j}_B$. Consequently for  $2\times 2$ sub-matrix
\begin{eqnarray}
\left(\begin{array}{cc} \rho_{ii} & \rho_{i(n+1-i)}\\ \rho_{(n+1-i)i} & \rho_{(n+1-i)(n+1-i)} \end{array}\right),
\end{eqnarray}
the eigenstates are  Schmidt decompositions in their own form. More importantly since  there is no overlap between different sub-matrices,  then the "nearest" $\sigma$ would be the probabilistic combination of the bases of all sub-matrices. We work out two important examples to display the validity of statement

\emph{Example 1.} Consider the  Werner state
\begin{eqnarray}\label{wernerstate}
\rho_{W}=\tfrac{1-r}{4}I+ r\ket{\psi^+}\bra{\psi^+},
\end{eqnarray}
in which $r\in [0,1]$ and $\ket{\psi^+}=\tfrac{1}{\sqrt{2}}(\ket{10}+\ket{01})$. It is obvious that $\rho_{W}$ has a "X" form on the basis $\left\{\ket{11}, \ket{10}, \ket{01}, \ket{00}\right\}$
\begin{eqnarray}
\rho_{W}=\frac{1}{4}\left(\begin{array}{cccc}
1-r&0&0&0\\
0&1+r&2r&0\\
0&2r&1+r&0\\
0&0&0&1-r
\end{array}\right),
\end{eqnarray}
which can be decomposed into two sub-matrices
\begin{eqnarray}
\rho_1=\tfrac{1}{4}\left(\begin{array}{cc} 1-r&0\\ 0&1-r \end{array}\right);
\rho_2=\tfrac{1}{4}\left(\begin{array}{cc} 1+r&2r\\ 2r&1+r \end{array}\right),
\end{eqnarray}
defined on the bases $\left\{\ket{11}, \ket{00}\right\}$ and $\left\{ \ket{10}, \ket{01}\right\}$ respectively.
Then there are four eigenstates
\begin{eqnarray}
\ket{1}&=&\ket{11};\ket{2}=\ket{00}\nonumber\\
\ket{3}&=&\frac{1}{\sqrt{2}}(\ket{10}-\ket{01})\nonumber\\
\ket{4}&=&\frac{1}{\sqrt{2}}(\ket{10}+\ket{01})
\end{eqnarray}
of which the Schmidt bases are $\{\ket{11}, \ket{00}\}$ and $\{\ket{10}, \ket{01}\}$ respectively.

By Eq.\eqref{bm}, one obtains
\begin{eqnarray}
\max\text{Tr} \left[\sqrt{\rho_W}\sqrt{\sigma}\right]=2\max\sqrt{c_0^2+c_1^2}
\end{eqnarray}
in which $c_0=\left(3\sqrt{1-r}+\sqrt{1+3r}\right)/8$, $c_1=\cos\Omega\left(\sqrt{1-r}-\sqrt{1+3r}\right)/8$ and $\cos \Omega=\cos\theta_1 \cos\theta_2- \sin\theta_1\sin\theta_2 \cos(\phi_1-\phi_2)$. Then one has $D^{H}=1- \tfrac{1}{2}\sqrt{3-r+\sqrt{(1-r)(1+3r)}}$ when $\cos\Omega=\pm 1$, which can occur, for example when $\theta_1=\theta_2=0$ and $\phi_1=\phi_2$. The corresponding "nearest" $\sigma$ has the form
\begin{eqnarray}
\sigma&=&\tfrac{1}{2}\tfrac{1+r+\sqrt{(1-r)(1+3r)}}{3-r+\sqrt{(1-r)(1+3r)}}\left(\ket{10}\bra{10}+\ket{01}\bra{01}\right)\nonumber\\
&&+\tfrac{1-r}{3-r+\sqrt{(1-r)(1+3r)}}\left(\ket{11}\bra{11}+\ket{00}\bra{00}\right),
\end{eqnarray}
which just is a mixed combination of the Schmidt bases of $\rho_1$ and $\rho_2$. It should be pointed out the the choice of  $\theta_{1(2)}$ and $\phi_{1(2)}$ is not unique.

\emph{Example 2.} Consider the Bell-diagonal state
\begin{eqnarray}
\beta^{ab}&=&\lambda_1\ket{\Psi^+}\bra{\Psi^+}+\lambda_2\ket{\Psi^-}\bra{\Psi^-}\nonumber \\ &&+\lambda_3\ket{\Phi^+}\bra{\Phi^+}+\lambda_4\ket{\Phi^-}\bra{\Phi^-},
\end{eqnarray}
in which $\ket{\Psi^{\pm}}=\tfrac{1}{\sqrt{2}}(\ket{00}\pm\ket{11})$, $\ket{\Phi^{\pm}}=\tfrac{1}{\sqrt{2}}(\ket{01}\pm\ket{10})$. In matrix,
\begin{eqnarray}
\beta^{ab}=\frac{1}{2}\left(\begin{array}{cccc}
\lambda_1+\lambda_2&0&0&\lambda_1-\lambda_2\\
0&\lambda_3+\lambda_4&\lambda_3-\lambda_4&0\\
0&\lambda_3-\lambda_4&\lambda_3+\lambda_4&0\\
\lambda_1-\lambda_2&0&0&\lambda_1+\lambda_2
\end{array}\right),
\end{eqnarray}
on local basis $\left\{\ket{11}, \ket{10}, \ket{01}, \ket{00}\right\}$. Or equivalently
\begin{eqnarray}
\beta^{ab}=\frac{1}{2}\left(\begin{array}{cccc}
\lambda_1+\lambda_3&0&0&\lambda_1-\lambda_3\\
0&\lambda_2+\lambda_4&\lambda_2-\lambda_4&0\\
0&\lambda_2-\lambda_4&\lambda_2+\lambda_4&0\\
\lambda_1-\lambda_3&0&0&\lambda_1+\lambda_3
\end{array}\right),
\end{eqnarray}
on local basis $\left\{\ket{1_x1_x}, \ket{1_x0_x}, \ket{0_x1_x}, \ket{0_x0_x}\right\}$.

Then by Eq.\eqref{bm}
\begin{eqnarray}
\max \text{Tr} \left[ \sqrt{\beta^{ab}} \sqrt{\sigma}\right]
=\tfrac{1}{4} \max \sqrt{2(c_1^2+c_2^2)},
\end{eqnarray}
in which
\begin{eqnarray}
&c_1=h+\lambda_1\cos\Omega_1^+ +\lambda_2\cos\Omega_1^- -\lambda_3\cos\Omega_2^- -\lambda_4\cos\Omega_2^+;\nonumber\\
&c_2=h-\lambda_1\cos\Omega_1^+ -\lambda_2\cos\Omega_1^- +\lambda_3\cos\Omega_2^- +\lambda_4\cos\Omega_2^+;\nonumber\\
&\cos\Omega_1^{\pm}=\cos\theta_1\cos\theta_2\pm\sin\theta_1\sin\theta_2\cos(\phi_1+\phi_2)\nonumber\\
&\cos\Omega_2^{\pm}=\cos\theta_1\cos\theta_2\pm\sin\theta_1\sin\theta_2\cos(\phi_1-\phi_2)\nonumber\\
&h=\sqrt{\lambda_1}+\sqrt{\lambda_2}+\sqrt{\lambda_3}+\sqrt{\lambda_4}.\nonumber
\end{eqnarray}
It is not difficult to find that the extremum happens when $\sin(\phi_1-\phi_2)=\sin(\phi_1+\phi_2)=0$. Then dependent on $\cos(\phi_1-\phi_2)=\pm 1$ and  $\cos(\phi_1+\phi_2)=\pm 1$, one has
\begin{eqnarray}
\max \text{Tr} \left[ \sqrt{\beta^{ab}} \sqrt{\sigma}\right]
=\tfrac{1}{2}\sqrt{h^2+\max\{d^2_1, d^2_2, d^2_3\}},
\end{eqnarray}
in which
\begin{eqnarray*}
d_1&=&\sqrt{\lambda_1}-\sqrt{\lambda_2}+\sqrt{\lambda_3}-\sqrt{\lambda_4};\nonumber\\
d_2&=&-\sqrt{\lambda_1}+\sqrt{\lambda_2}+\sqrt{\lambda_3}-\sqrt{\lambda_4};\nonumber\\
d_3&=&\sqrt{\lambda_1}+\sqrt{\lambda_2}-\sqrt{\lambda_3}-\sqrt{\lambda_4}.
\end{eqnarray*}
$\sigma$ can be obtained by Eq. \eqref{bmp}; When $\theta_1=\theta_2=\pi/2, \phi_1=\phi_2=0$, $D^H(\beta^{ab})=1- \tfrac{1}{2}\sqrt{h^2+d^2_1}$. Then
\begin{eqnarray}
\sigma_1&=&\tfrac{1}{4(h^2+d_1^2)}\left[(h+d_1)^2\left(\ket{1_x 1_x}\bra{1_x 1_x}+\ket{0_x 0_x}\bra{0_x 0_x}\right)\right.\nonumber \\
&&\left.+(h-d_1)^2\left(\ket{1_x 0_x}\bra{1_x 0_x}+\ket{0_x 1_x}\bra{0_x 1_x}\right)\right].
\end{eqnarray}
When $\theta_1=\theta_2=\pi/2= \phi_1=\phi_2$, $D^H(\beta^{ab})=1- \tfrac{1}{2}\sqrt{h^2+d^2_2}$. Then
\begin{eqnarray}
\sigma_2&=&\tfrac{1}{4(h^2+d_2^2)}\left[(h+d_2)^2\left(\ket{1_x 1_x}\bra{1_x 1_x}+\ket{0_x 0_x}\bra{0_x 0_x}\right)\right.\nonumber \\
&&\left.+(h-d_2)^2\left(\ket{1_x 0_x}\bra{1_x 0_x}+\ket{0_x 1_x}\bra{0_x 1_x}\right)\right].
\end{eqnarray}
As for $D^H(\beta^{ab})=1- \tfrac{1}{2}\sqrt{h^2+d^2_3}$, there are two cases; When $\theta_1=\theta_2=0, \pi$,
\begin{eqnarray}
\sigma_3&=&\tfrac{1}{4(h^2+d_3^2)}\left[(h+d_3)^2\left(\ket{1_x 1_x}\bra{1_x 1_x}+\ket{0_x 0_x}\bra{0_x 0_x}\right)\right.\nonumber \\
&&\left.+(h-d_3)^2\left(\ket{1_x 0_x}\bra{1_x 0_x}+\ket{0_x 1_x}\bra{0_x 1_x}\right)\right].
\end{eqnarray}
When $\theta_1=0, \theta_2=\pi$ or $\theta_1=\pi, \theta_2=0$,
\begin{eqnarray}
\sigma_3&=&\tfrac{1}{4(h^2+d_3^2)}\left[(h-d_3)^2\left(\ket{1_x 1_x}\bra{1_x 1_x}+\ket{0_x 0_x}\bra{0_x 0_x}\right)\right.\nonumber \\
&&\left.+(h+d_3)^2\left(\ket{1_x 0_x}\bra{1_x 0_x}+\ket{0_x 1_x}\bra{0_x 1_x}\right)\right].
\end{eqnarray}
In both of cases, $\sin(\phi_1-\phi_2)=\sin(\phi_1+\phi_2)=0$ should be satisfied. It is obvious that $\sigma$ is detemined completely by the lcoal bases.

Finally we should point out that our result is compatible  with $D_H(\beta^{ab})$, defined by Eq. (10) in Ref. \cite{chang13}. Moreover $D^H(\beta^{ab})$ is less than $D_H(\beta^{ab})$. This compatibility display strongly the validity of our statement.

Then we obtain the second conjecture
\begin{theorem}\label{c2}
For  $X$-type density operator defined by Eq. \eqref{xstate}, the "nearest" completely classical state $\sigma$ is necessarily the mixed combination of the bases $\ket{i}_A\ket{j}_B$ with the joint probability distribution decided by Eq.\eqref{bmp}.
\end{theorem}

\section{Multipartite state: Symmetric case}

Since the absence of Schmidt decomposition in multipartite case, we focus on the  states of symmetry in this section, which is defined as the invariance under the permutation or  translation of  single-party states. Through several examples  we would demonstrate that the "nearest" $\sigma$ necessarily displays the same invariance as that of the measured state.

\subsection{3-qubit case}
\emph{-GHZ state-} $\ket{\text{GHZ}}=\tfrac{1}{\sqrt{2}}(\ket{111}+\ket{000})$ is obviously invariant under permutation. Then
\begin{eqnarray}
&&\sum_n\left|\inp{\text{GHZ}}{\sigma_n}\right|^4\nonumber\\
&=&\frac{1}{8}\left( 1 + \cos^2\theta_1 \cos^2\theta_2+ \cos^2\theta_1 \cos^2\theta_3+ \cos^2\theta_2 \cos^2\theta_3\right.\nonumber\\
&&\left.+\cos^2(\phi_1+\phi_2+\phi_3)\prod_{i=1}^3\sin^2\theta_i\right),
\end{eqnarray}
of which the maximal value  is determined by relations
\begin{eqnarray}
&\frac{\partial \sum_n\left|\inp{\text{GHZ}}{\sigma_n}\right|^4} {\partial \phi_i}=0\Rightarrow \sin(\phi_1+\phi_2+\phi_3)=0\nonumber\\
&\frac{\partial \sum_n\left|\inp{\text{GHZ}}{\sigma_n}\right|^4} {\partial \theta_i}=0\Rightarrow
\sin 2\theta_i=0.
\end{eqnarray}
The values for $\phi_i$ and $\theta_i$ are not unique; for  $\phi_1=\phi_2=\phi_3=0$ and $\theta_1=\theta_2=\theta_3=0$,  $D^{H}=1-\tfrac{1}{\sqrt{2}}$. So $\sigma=\tfrac{1}{2}\left(\ket{1}\bra{1}^{\otimes 3}+\ket{0}\bra{0}^{\otimes 3}\right)$, which obviously  is  invariant under permutation.

\emph{-$W$ state-} $\ket{W}=\tfrac{1}{\sqrt{3}}(\ket{100}+\ket{010}+\ket{001})$ is also invariant by permutation. By explicit calculation, one finds
\begin{eqnarray}
&&\sum_n\left|\inp{W}{\sigma_n}\right|^4\nonumber\\
&=&\tfrac{1}{72}\left(9+a^2+b^2+c^2+d^2+\sum_{i=1}^3\cos^2\theta_i\right),
\end{eqnarray}
in which
\begin{eqnarray}
a&=&3\cos\theta_1\cos\theta_2\cos\theta_3-2\cos\delta\phi_1\sin\theta_1\sin\theta_2\cos\theta_3\nonumber\\
&&-2\cos\delta\phi_2\sin\theta_1\cos\theta_2\sin\theta_3\nonumber\\
&&-2\cos\delta\phi_3\cos\theta_1\sin\theta_2\sin\theta_3,\nonumber\\
b&=&\cos\theta_1\cos\theta_2-2\cos\delta\phi_1\sin\theta_1\sin\theta_2,\nonumber\\
c&=&\cos\theta_1\cos\theta_3-2\cos\delta\phi_2\sin\theta_1\sin\theta_3,\nonumber\\
d&=&\cos\theta_2\cos\theta_3-2\cos\delta\phi_3\sin\theta_2\sin\theta_3,\nonumber\\
\delta\phi_1&=&\phi_1-\phi_2;\delta\phi_2=\phi_1-\phi_3;\delta\phi_3=\phi_2-\phi_3.
\end{eqnarray}
It is not difficult to find that the extremum appears when $\sin\delta \phi_i=0$. As for $\theta_i$, one can find by thorough calculation that $\max\sum_n\left|\inp{W}{\sigma_n}\right|^4=1/3$  when $\theta_i=0, \pi(i,j=1,2,3)$. Thus  $D^{H}=1-\tfrac{1}{\sqrt{3}}$ and
\begin{eqnarray}
\sigma&=&\tfrac{1}{3}\left(\ket{1}\bra{1}\otimes\ket{0}\bra{0}\otimes\ket{0}\bra{0}+\ket{0}\bra{0}\otimes\ket{1}\bra{1}\otimes\ket{0}\bra{0} \right.\nonumber\\
&&\left.+\ket{0}\bra{0}\otimes\ket{0}\bra{0}\otimes\ket{1}\bra{1}\right),
\end{eqnarray}
when, for instance, $\theta_i=\phi_i=0$. Obviously $\sigma$ is permutationally invariant too.

\subsection{4-qubit case}
As for 4-qubit states, there exist a different invariance  from the  permutational, termed as translational invariance. Its meaning is similar to that in solid systems;  the difference is that it refer to single-party state in Hilbert space in this place, instead of single particle in real lattice configuration \cite{cui}. We will display by two examples that the "nearest" $\sigma$ is necessarily  translationally invariant too.

\emph{-$\ket{\text{GHZ}_1}_4$ state-}, which is defined as $\ket{\text{GHZ}_1}_4=\tfrac{1}{\sqrt{2}}(\ket{1010}+\ket{0101})$. It is obvious that the state is actually constructed by cyclic permutation of $1010$, which is named as \emph{cyclic unit}. It is not difficult to find
\begin{eqnarray}
&&\sum_n\left|_4\inp{\text{GHZ}_1}{\sigma_n}\right|^4\nonumber\\
&=&\tfrac{1}{16}\left(1+a^2+b^2+c^2+d^2+e^2+f^2+g^2\right),
\end{eqnarray}
in which
\begin{eqnarray*}
a&=&\prod_{i=1}^4 \cos\theta_i+\cos(\phi_1-\phi_2+\phi_3-\phi_4)\prod_{i=1}^4\sin\theta_i,\nonumber\\
b&=&\cos\theta_1\cos\theta_2; c=\cos\theta_1\cos\theta_3;d=\cos\theta_1\cos\theta_4,\nonumber\\
e&=&\cos\theta_2\cos\theta_3;f=\cos\theta_2\cos\theta_4;g=\cos\theta_3\cos\theta_4,
\end{eqnarray*}
which has maximal value $1/2$ and then  $D^H=1-1/\sqrt{2}$ when $\cos\theta_i=\pm 1(i=1,2,3,4)$ and  $\sin(\phi_1-\phi_2+\phi_3-\phi_4)=0$. Consequently  $\sigma=\tfrac{1}{2}(\ket{1010}\bra{1010}+\ket{0101}\bra{0101})$ by setting $\phi=0$, which is also translationally invariant.

\emph{-$\ket{W_2}_4 state$-}, which is defined as
\begin{eqnarray}
\ket{W_2}_4=\tfrac{1}{2}\left(\ket{1100}+\ket{0110}+\ket{0011}+\ket{1001}\right).
\end{eqnarray}
The state is actually constructed by cyclic unit $1100$. Moreover it is bi-seperable, $\ket{W_2}_4=\tfrac{1}{\sqrt{2}}\left(\ket{10}+\ket{01}\right)_{13}\otimes \tfrac{1}{\sqrt{2}}\left(\ket{10}+\ket{01}\right)_{24}$.  Thus the "nearest" $\sigma$ can  also be factorized into two parts, i.e. $\sigma=\sigma_{13}\otimes\sigma_{24}$, in which $\sigma_{13}$ and $\sigma_{24}$ are the "nearest" completely classical states for $\tfrac{1}{\sqrt{2}}\left(\ket{10}+\ket{01}\right)_{13}$ and $ \tfrac{1}{\sqrt{2}}\left(\ket{10}+\ket{01}\right)_{24}$ respectively. By Conjecture \ref{c1}, one can obtain $D^H=1/2$ and
\begin{eqnarray}
\sigma&=&\tfrac{1}{2}\left(\ket{10}\bra{10}+\ket{01}\bra{01}\right)_{13}\otimes\tfrac{1}{2}\left(\ket{10}\bra{10}+\ket{01}\bra{01}\right)_{24}\nonumber\\
&=&\tfrac{1}{4}\left(\ket{1100}\bra{1100}+\ket{0110}\bra{0110}+\ket{0011}\bra{0011}\right.\nonumber\\
&&\left.+\ket{1001}\bra{1001}\right),
\end{eqnarray}
which is obviously translationally invariant.

\subsection{A short discussion}
We can obtain the third conjecture
\begin{theorem}\label{c3}
For multipartite state with permutational or translational invariance, the "nearest" $\sigma$ necessarily has the same invariance.
\end{theorem}
It should be pointed out that the form of  $\sigma$ cannot be obtained directly from the measured state in general.  For example, we try to find the $\sigma$ for Dicke state $\ket{4, 2}=\tfrac{1}{\sqrt{6}}\sum_{\text{perm.}}\ket{1100}=\tfrac{1}{\sqrt{3}}\ket{\text{GHZ}_1}_4 +\sqrt{\tfrac{2}{3}}\ket{W_2}_4$. By explicit calculation, one obtain
\begin{eqnarray}
&&\sum_n\left|\inp{4,2}{\sigma_n}\right|^4\nonumber\\
&=&\frac{217-96 \cos (2 \theta )+108 \cos (4 \theta )+27 \cos (8 \theta )}{1536},
\end{eqnarray}
which has maximal value when $\theta=\pi/2$. Then $D^H\approx 0.46$ and
\begin{eqnarray}
\sigma&=&0.482\left(\ket{1_x}\bra{1_x}^{\otimes 4}+\ket{0_x}\bra{0_x}^{\otimes 4}\right)\nonumber\\
&&+0.006\left(\sum_{\text{perm}}\ket{1_x}\bra{1_x}^{\otimes 2}\ket{0_x}\bra{0_x}^{\otimes 2}\right),
\end{eqnarray}
in which $\sum_{\text{perm}}$ denotes the  permutations of the density operators with two qubits being state $\ket{1_x}$ and the other two being state $\ket{0_x}$.

\section{$D^H$ and quantum phase transition in many-body system}
In this section, we show that $D^H$ can also mark the quantum phase transition in many-body systems. For clarity, this discussion focuses on two popular models, Lipkin-Meshkov-Glick (LMG) \cite{lmg} and Dicke models \cite{dicke}, of which the ground states can be determined analytically.

\subsection{Lipkin-Meshkov-Glick model}

The LMG model describes a set of spin-half particles coupled to all others with an interaction independent of the position and the nature of the elements. The Hamiltonian can be written as
\begin{equation}\label{lmg}
H= - \frac{\lambda}{N}(S^2_x + \gamma S^2_y) - h_z S_z,
\end{equation}
in which $S_{\alpha}=\sum_{i=1}^{N}\sigma^i_{\alpha}/2 (\alpha=x, y, z)$ and the $\sigma_{\alpha}$ denotes  the Pauli operator, and $N$ is the total particle number in this system. The prefactor $1/N$ is essential to ensure the convergence of the free energy per spin in the thermodynamic limit. It is known that there is a second-order transition at $h=h_z/|\lambda|=1$ for the ferromagnetic case ($\lambda>0$) and a first-order one
at $h=0$ for the antiferromagnetic case ($\lambda<0$) \cite{botet, vidal}. The following discussion is divided into two parts by $\gamma=1$ or not.

\emph{-$\gamma=1$-} In this case the model can be solved exactly; the eigenstate is  $\ket{N/2, n}$, in which $N/2,n$ denote the quantum numbers of the total angular momentum $S^2$ and  $S_z$, and the corresponding eigenenergy is $E_n=-\tfrac{\lambda}{2}(\tfrac{N}{2}+1)+\tfrac{\lambda}{N}n^2-h_z n (\hbar=1)$. For $\lambda>0$, the minimal value of $E_n$ appears when $n=\left[\tfrac{h_z}{\lambda}\tfrac{N}{2}\right]$. Then the ground state is $\ket{N/2, N/2}$ for $h_z/\lambda>1$. The state is Dicke state $\ket{N, N}_{\text{Dicke}}$, for which obvioulsy $D^H=0$. As for $h_z/\lambda<1$, the ground state is $\ket{N/2, \left[\tfrac{h_z}{\lambda}\tfrac{N}{2}\right]}$, which can be rewritten as Dicke state\begin{eqnarray}
\ket{N, m}_{\text{Dicke}}=\tfrac{1}{\sqrt{^N C_m}}\sum_{\text{perm}}\ket{\underbrace{1\cdots 1 }_{m} \underbrace{0\cdots 0}_{N-m}},
\end{eqnarray}
in which $m=\tfrac{N}{2}+\left[\tfrac{h_z}{\lambda}\tfrac{N}{2}\right]$.

\begin{figure}[t!!]
\center
\includegraphics[width=\columnwidth]{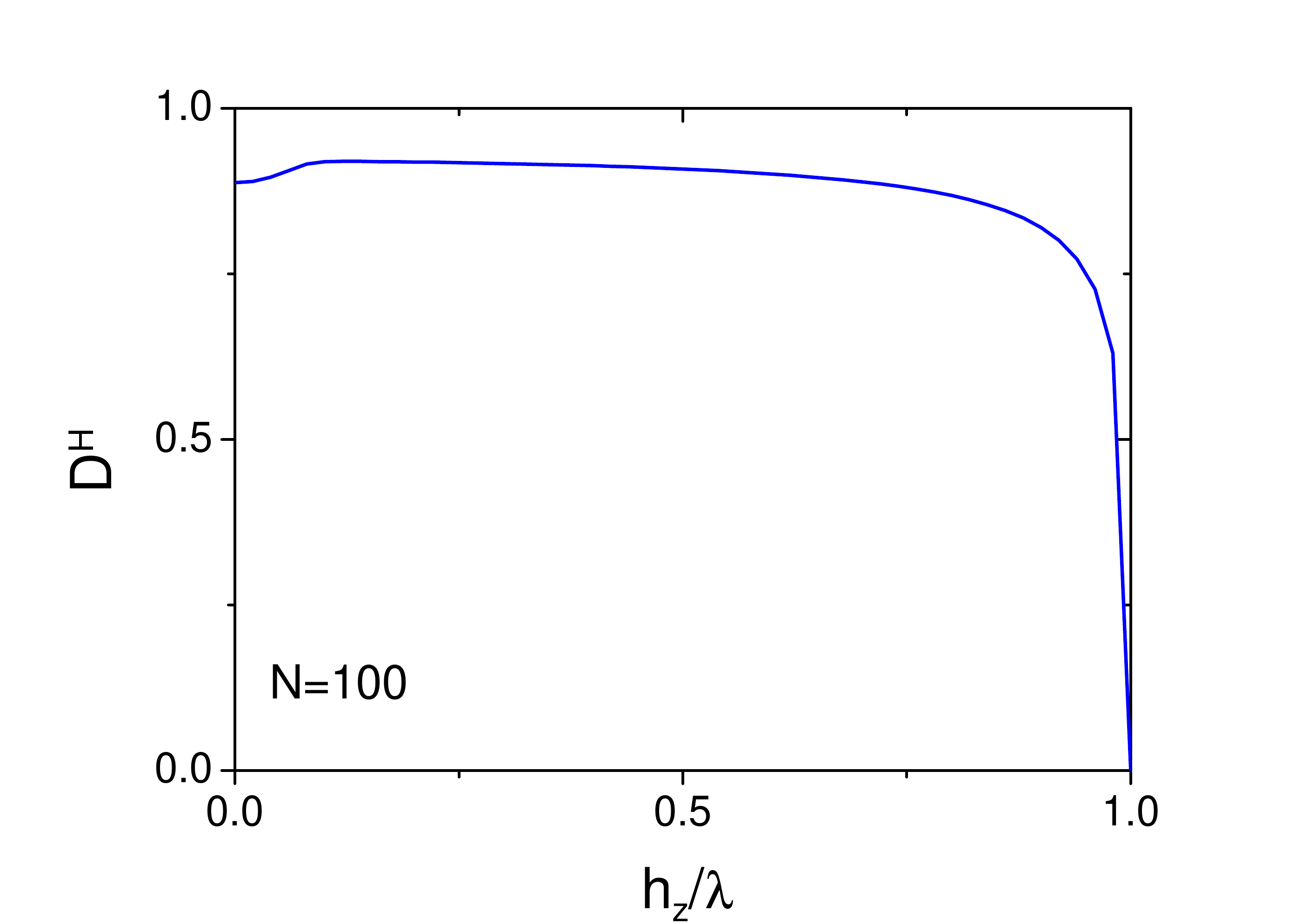}
\caption{(Color online) $D^H$ of the ground state for $\gamma=1$ and $\lambda>0$ in LMG model.}
\label{fig:lmg1}
\end{figure}

It is obvious that the ground state is permutationally invariant. Thus with respect of Dicke stat $\ket{N, m}_{\text{Dicke}}$  $\sigma$ can be written directly as by Conjecture \ref{c3}
\begin{eqnarray}\label{msigma}
\sigma&=&\sum_{m=0}^{N}p_m \sigma_m ;\nonumber \\
\sigma_m&=&\sum_{\text{perm}}\ket{\underbrace{+\cdots +}_{m} \underbrace{-\cdots -}_{N-m}}\bra{+ \cdots + -\cdots -},
\end{eqnarray}
in which $\ket{+}=\cos \frac{\theta}{2} \ket{1}+\exp(-i\phi)\sin \frac{\theta}{2}\ket{0}$ and $\ket{-}=\exp(i\phi)\sin \frac{\theta}{2}\ket{1}-\cos \frac{\theta}{2}\ket{0}$ with $\theta\in[0, \pi]$ and $\phi\in[0, 2\pi)$.  $p_m$ can be determined by Eq. \eqref{bp}. It should be pointed out that because of permutational invariance the probability $p_m$ is same for the local states $\ket{\underbrace{+\cdots +}_{m} \underbrace{-\cdots -}_{N-m}}$ with the same $m$. By numerical evaluation one has
\begin{eqnarray}
D^H=\left\{ \begin{array}{cc} 0, &  h_z/\lambda>1; \\
>0, & 0<h_z/\lambda<1.\end{array} \right.,
\end{eqnarray}
which is  plotted for $0<h_z/\lambda<1$ in Fig.\ref{fig:lmg1}.

As for $\lambda<0$, the minimal value of $E_n$ appears when $n=- \left[\tfrac{h_z}{|\lambda |}\tfrac{N}{2}\right]$. Then the ground state is $\ket{N/2, N/2}$ for $h_z>0$ and $\ket{N/2, -N/2}$ for $h_z<0$ in angular moment picture,  which are Dicke states $\ket{N, N}_{\text{Dicke}}$ and $\ket{N, 0}_{\text{Dicke}}$ respectively.  Thus $D^H=0$.

\emph{-$\gamma\in[0,1)$-} The ground state is  \cite{cui06}
\begin{eqnarray}
\label{g}
\ket{g}&=&\frac{1}{c}\sum_{n=0}^{[N/2]}(-1)^n\sqrt{\tfrac{(2n-1)!!}{2n!!}}\tanh^nx\ket{N,N-2n}_{\text{Dicke}}\nonumber\\
c^2&=&\sum_{n=0}^{[N/2]}(-1)^n\tfrac{(2n-1)!!}{2n!!}\tanh^{2n}x,
\end{eqnarray}
in which
\begin{eqnarray}\label{ft}
\tanh 2x=\begin{cases}-\frac{1-\gamma}{2h_z-1-\gamma},&
h_z>1\\-\frac{h_z^2-\gamma}{2-h_z^2-\gamma}, & 0\leq h_z<1\end{cases}.
\end{eqnarray}
for $\lambda>0$ and
\begin{eqnarray}\label{aft}
\tanh 2x=\frac{1-\gamma}{1+\gamma+2|h_z|}.
\end{eqnarray}
for $\lambda<0$.

\begin{figure}[t!!]
\center
\includegraphics[width=\columnwidth]{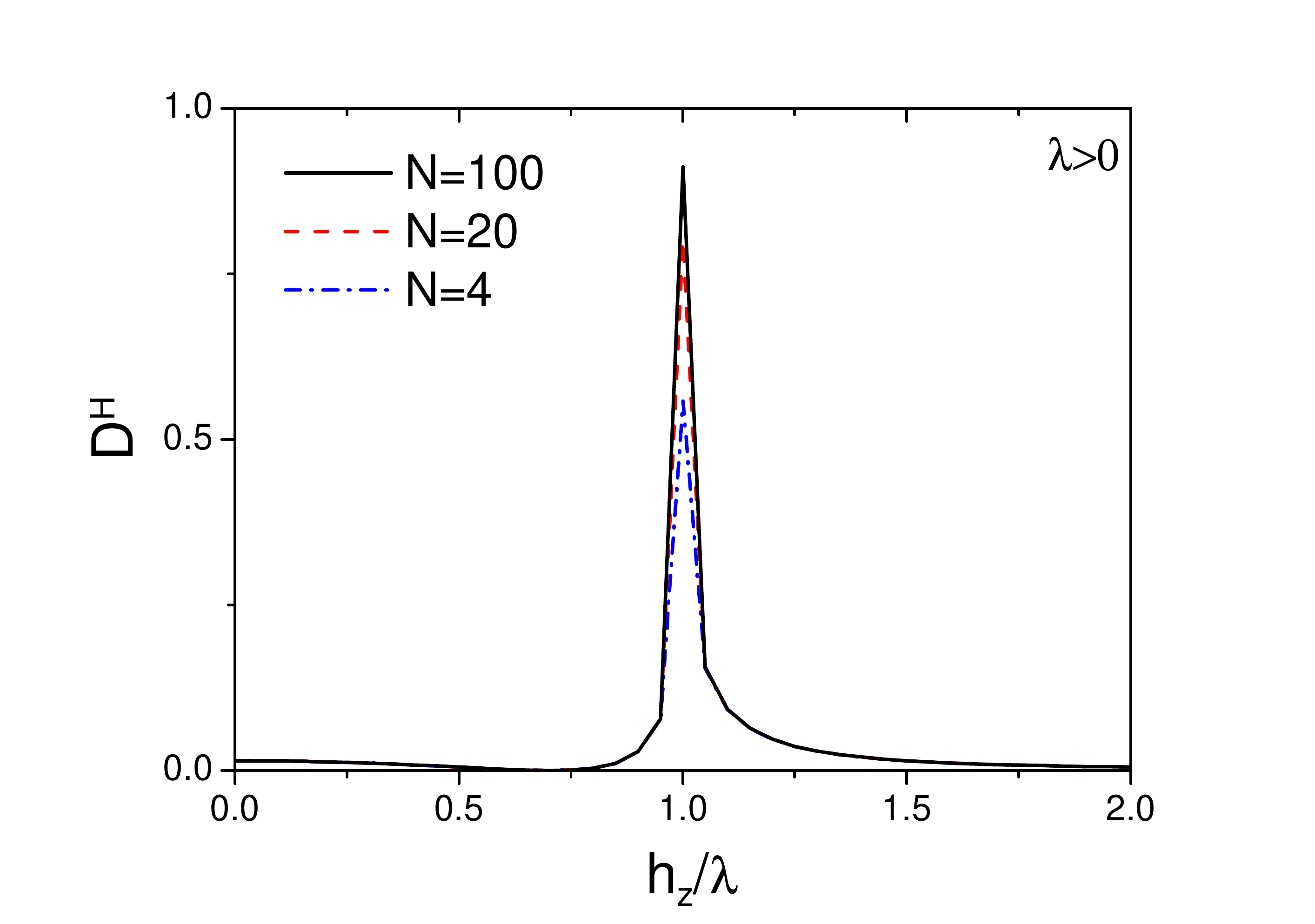}
\includegraphics[width=\columnwidth]{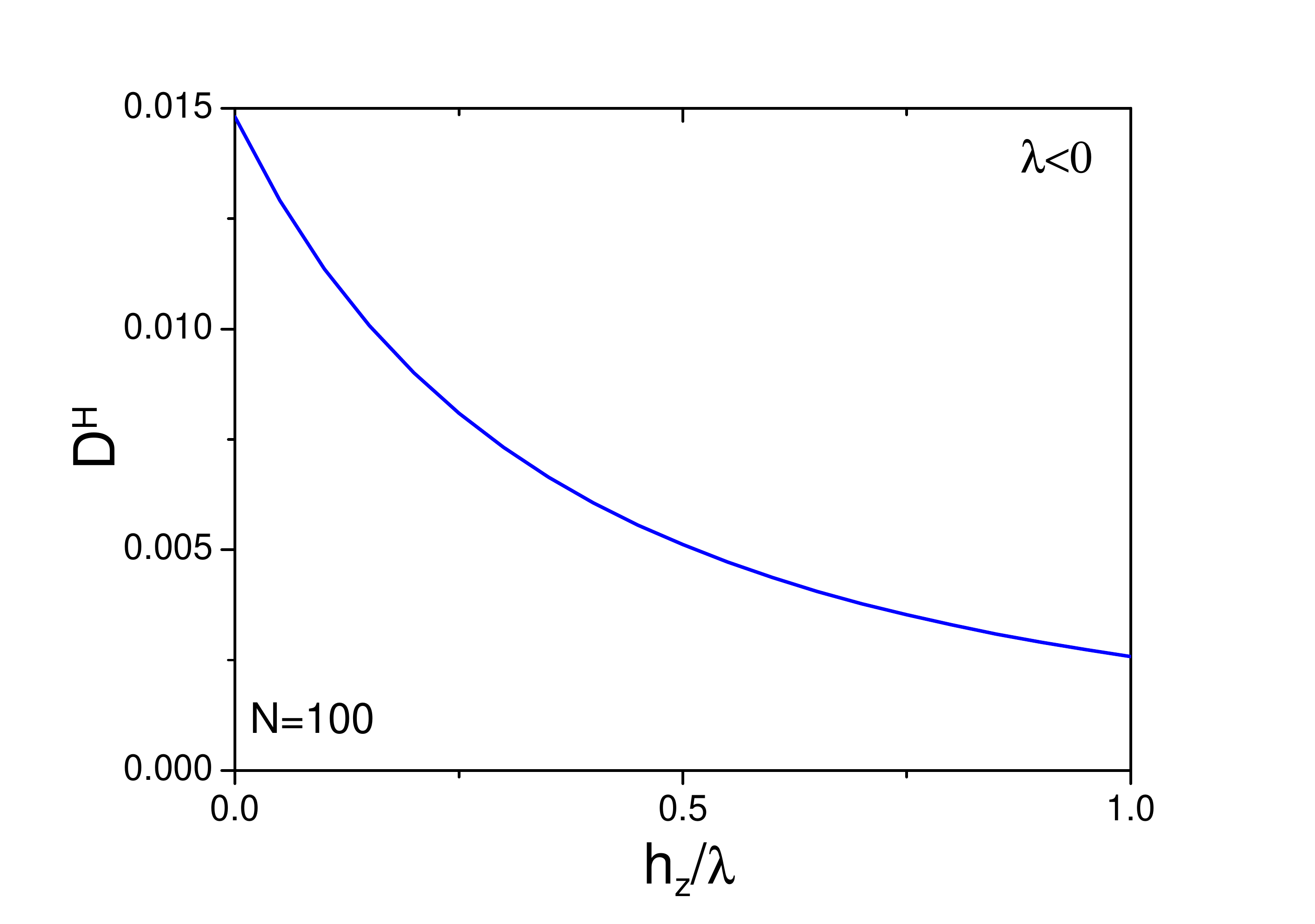}
\caption{(Color online) $D^H$ of the ground state for $\gamma=1/2$ in LMG model.}
\label{fig:lmg23}
\end{figure}

In this case $\sigma$ has the same form as Eq.\eqref{msigma}. In Fig.\ref{fig:lmg23}, $D^H$ is plotted, in which the critical points can be identified, $h_z/\lambda =1$ for $\lambda>0$ and $h_z/\lambda =0$ for $\lambda<0$.

\begin{figure}[t!!]
\center
\includegraphics[width=\columnwidth]{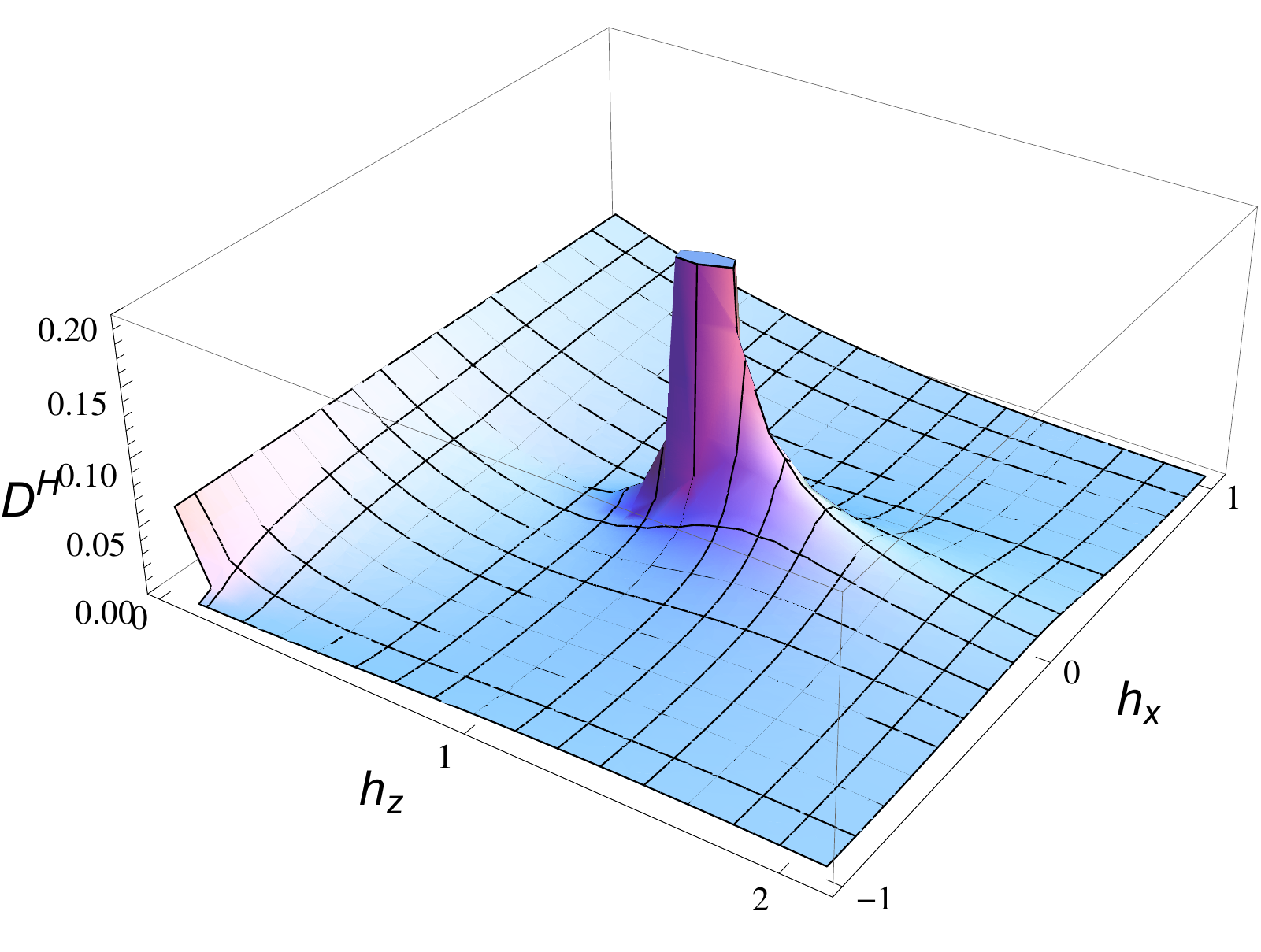}
\includegraphics[width=\columnwidth]{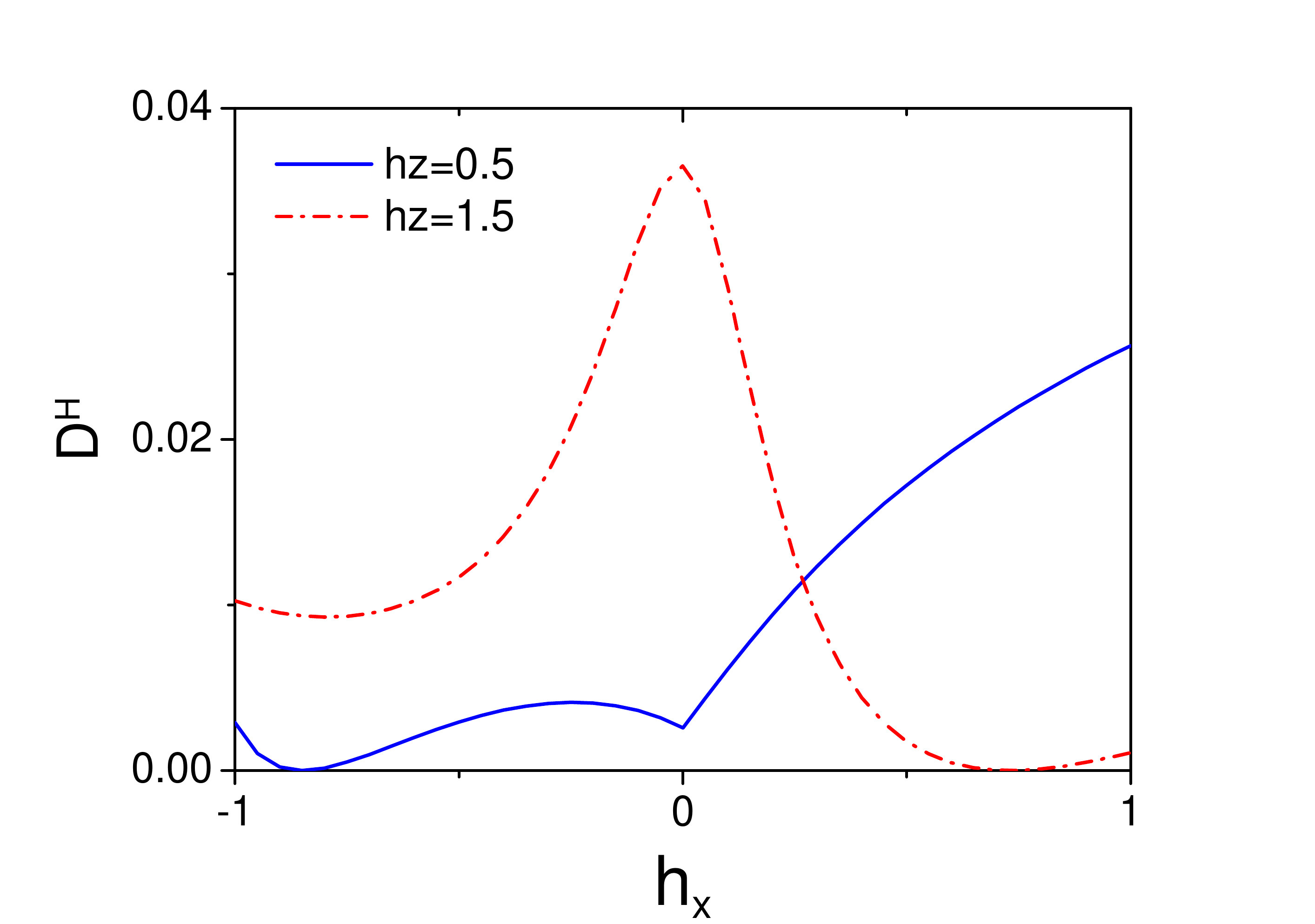}
\includegraphics[width=\columnwidth]{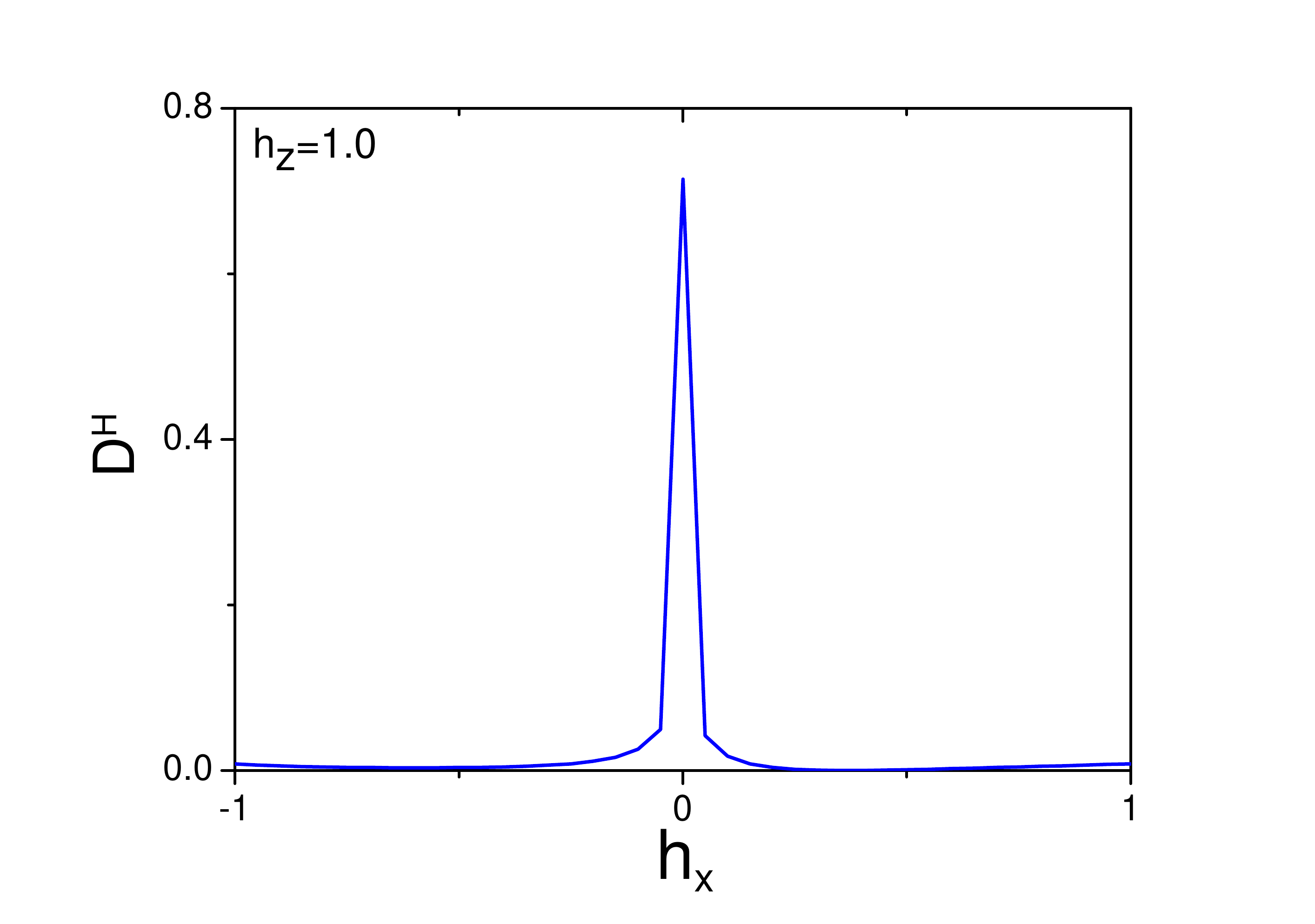}
\caption{(Color online) $D^H$ of the ground state for uniaxial model when $N=20$.}
\label{fig:lmg4}
\end{figure}

A generalization of LMG is the so-called uniaxial model,
\begin{eqnarray}
H = - \frac{1}{N}S^2_x - h_x S_x -h_z S_z.
\end{eqnarray}
The ground state has the same form to Eq.\eqref{g} with $\tanh 2x=\frac{2\Gamma(\lambda_0)}{\Delta(\lambda_0)}$, in which
\begin{eqnarray}
\Gamma(\lambda_0)&=&-\frac{1-5\lambda_0^2}{4}+h_x\frac{\lambda_0(2-\lambda_0^2)}{8(1-\lambda_0^2)^{3/2}}\nonumber\\
\Delta(\lambda_0)&=&h_z -
\frac{1-7\lambda_0^2}{2}+h_x\frac{\lambda_0(4-3\lambda_0^2)}{4(1-\lambda_0^2)^{3/2}},
\end{eqnarray}
and $\lambda_0$ is determined by the equation
\begin{equation}
\lambda_0 h_z - \frac{h_x(1-2\lambda_0^2)}{2\sqrt{1-\lambda_0^2}}-\lambda_0(1-2\lambda_0^2)=0.
\end{equation}
There are two critical points, $h_x=0$ for $h_z=1$, which corresponds to a second order quantum phase transition and $h_x=0$ for $0<h_z<1$, a first order one.  As shown in Fig. \ref{fig:lmg4}, $D^H$ can unambiguously manifest the appearance of critical points.

\subsection{Dicke model}
Dicke model \cite{dicke}is related to many fundamental issues in quantum optics, quantum mechanics and condensed matter physics, such as the coherent
spontaneous radiation\cite{hh73}, the dissipation of quantum system \cite{leggett87}, quantum chaos\cite{hakee} and atomic self-organization in a cavity\cite{bgbe09}. The multipartite entanglement in Dicke model has also been discussed \cite{multi-entanglement}. The Hamiltonian for single-model Dicke model reads
\begin{eqnarray}\label{dicke}
H&=&\omega a^{\dagger}a + \frac{\omega_0}{2}\sum_{i=1}^N\sigma_i^z +
\frac{\lambda}{\sqrt{N}}\sum_{i=1}^N
(\sigma^+_i+\sigma^-_i)(a^{\dagger}+a)\nonumber\\ &=& \omega_0 J_z +
\omega
a^{\dagger}a+\frac{\lambda}{\sqrt{N}}(a^{\dagger}+a)(J_++J_-),
\end{eqnarray}
where $J_z=\sum_{i=1}^N\sigma_i^z/2$ and $J_{\pm}=\sum_{i=1}^N\sigma_i^{\pm}$ are the collective angular momentum operators.  There are two distinct phases for ground state, normal phase and superradiant phase, separated by critical point $\lambda_c=\sqrt{\omega\omega_0}/2$.

\begin{figure}[t!!]
\center
\includegraphics[width=\columnwidth]{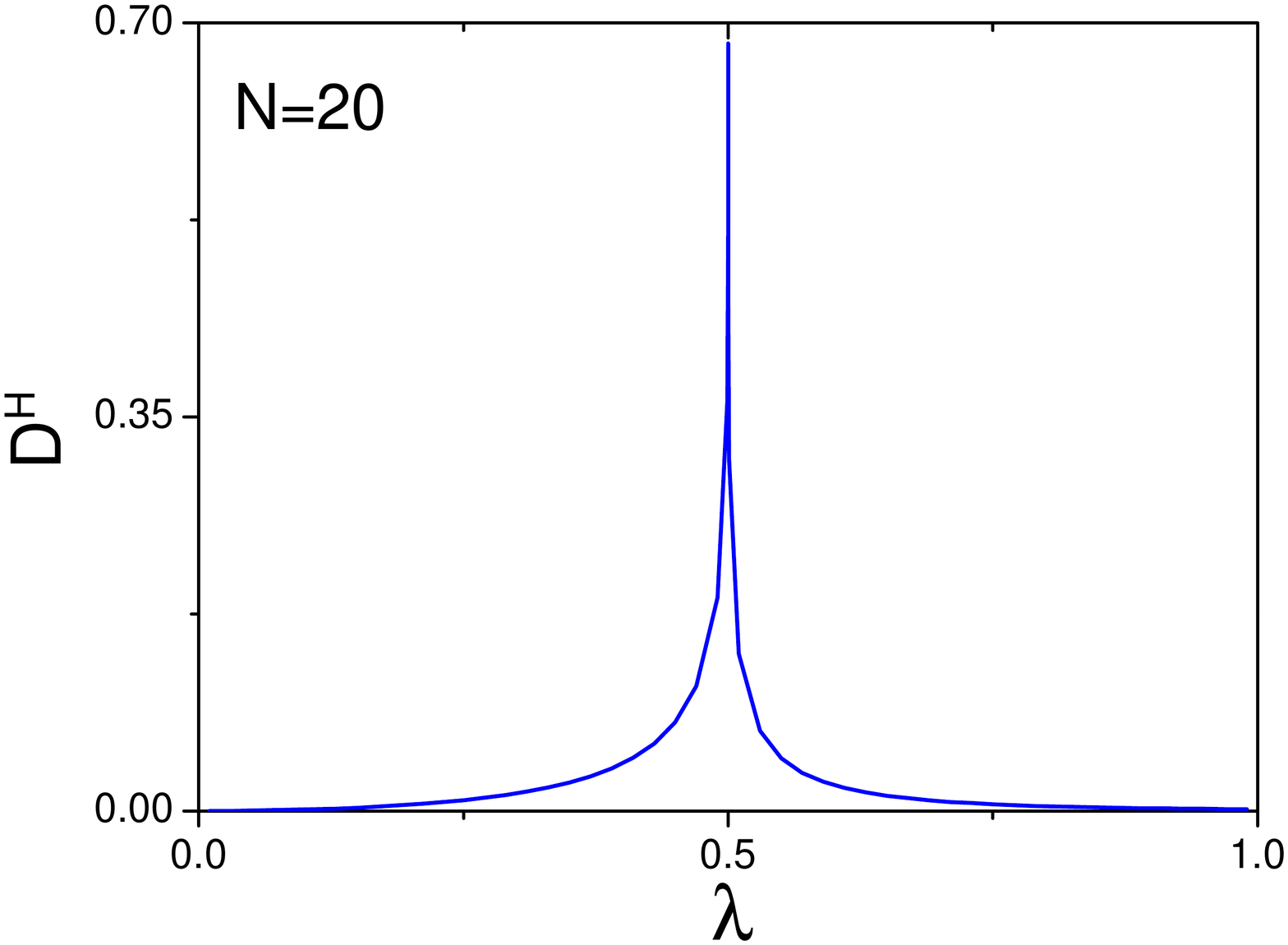}
\caption{(Color online) $D^H$ of the ground state for Dicke model when $N=20$, in which we has set $\omega=\omega_0=1$ and the critical point $\lambda_c=0.5$.}
\label{fig:dickemodel}
\end{figure}

By the method in Ref. \cite{eb03}, the reduce density operator of atom system for the ground state can be obtained analytically, which has the form
\begin{eqnarray}
\rho=\sum_k \lambda_k\ket{k}\bra{k},
\end{eqnarray}
in which $\ket{k}=\sum_{n_k=0}^N C_{n_k}\ket{N, n_k}_{\text{Dicke}}$. It is obvious that $\rho$ is invariant under permutation. Then $\sigma$ is Eq. \eqref{msigma}, and  $ D^H$ can be evaluated by numerical way, as shown in Fig. \ref{fig:dickemodel}. Obviously it clearly marks the appearance of quantum phase transition.

\section{Multilevel cases }
The conjectures in this article can be generalized directly into  multilevel case. In this section, we try to show the validity  in comparison with two exactly solved examples in Ref. \cite{chang13}.

\emph{Example 1. } $(m\times m)$-dimensional Werner state
\begin{eqnarray}
\omega^{ab}=\tfrac{m-x}{m^3 -m} \mathbf{I}^{ab}+\tfrac{mx-1}{m^3 -m}F,
\end{eqnarray}
in which $x\in[-1, 1]$ and $F=\sum_{k, l=1}^m \ket{kl}\bra{lk}$. In matrix, $\omega^{ab}$ is the direct sum of the following sub-matrices
\begin{eqnarray}
\omega_1^{ab}&=&\tfrac{1+x}{m^2 +m}\sum_{k=1}^m \ket{kk}\bra{kk} ;\nonumber\\
\omega_2^{ab}&=&\sum_{k \neq l}  \omega_{kl}^{ab},
\end{eqnarray}
in which
\begin{eqnarray}
\omega_{kl}^{ab}&=&\tfrac{m-x}{m^3 -m}\left(\ket{kl}\bra{kl}+\ket{lk}\bra{lk}\right)\nonumber \\
&&+\tfrac{mx-1}{m^3 -m}\left(\ket{kl}\bra{lk}+\ket{lk}\bra{kl}\right).
\end{eqnarray}
Thus $\omega^{ab}$ is actually "X" form, defined in Eq. \eqref{xstate}.

It is easy to find
\begin{eqnarray}
\sqrt{\omega^{ab}}&=&\sqrt{\tfrac{1+x}{m^2 +m}}\sum_{k=1}^m \ket{kk}\bra{kk} + \sqrt{\tfrac{1+x}{m^2 +m}}\sum_{k\neq l} \ket{1_{kl}}\bra{1_{kl}}\nonumber \\
&& +\sqrt{\tfrac{1-x}{m^2 -m}}\sum_{k\neq l} \ket{0_{kl}}\bra{0_{kl}},
\end{eqnarray}
in which $\ket{1_{kl}}=\tfrac{1}{\sqrt{2}}(\ket{kl}+\ket{lk})$ and $\ket{0_{kl}}=\tfrac{1}{\sqrt{2}}(\ket{kl}-\ket{lk})$. By Conjecture 2, the nearest neighbor $\sigma$ is
\begin{eqnarray}
\sigma=\sum_{k}p_k\ket{kk}\bra{kk} + \sum_{k\neq l} p_{kl} \ket{kl}\bra{kl},
\end{eqnarray}
in which $p_k$ and $p_{kl}$ is determined by Eq. \eqref{bmp}.  Thus by calculations,
\begin{eqnarray}
&&\max \text{Tr}\left[ \sqrt{\omega^{ab} }\sqrt{\sigma} \right]\nonumber \\
&=&\left\{ m \tfrac{1+x}{m+1} +\tfrac{m^2 - m}{4}\left[ \sqrt{\tfrac{1+x}{m^2 +m}}\ + \sqrt{\tfrac{1-x}{m^2 -m}}\ \right]^2  \right\}^{1/2}
\end{eqnarray}
Thus one has
\begin{eqnarray}
D^H(\omega^{ab})=1- \frac{1}{\sqrt{2}}\left[\frac{2+m+x}{m+1} + \sqrt{\frac{m-1}{m+1}} \sqrt{1-x^2} \right]^{1/2}.
\end{eqnarray}
Compared with the Eq. (15) in Ref. \cite{chang13}
\begin{eqnarray}
D_H(\omega^{ab})=1- \frac{1}{2}\left[\frac{2+m+x}{m+1} + \sqrt{\frac{m-1}{m+1}} \sqrt{1-x^2} \right],
\end{eqnarray}
it is not difficult to find that the two results are compatible and  $D^H(\omega^{ab}) < D_H(\omega^{ab})$.

\emph{Example 2.} $(m\times m)$-dimensional isotropic state
\begin{eqnarray}
\varsigma^{ab}= \frac{1- x}{m^2 -1}\mathbf{I}^{ab}+ \frac{mx^2 - 1}{m^2 -1}\ket{\Psi^+}\bra{\Psi^+},
\end{eqnarray}
in which $\ket{\Psi^+}=\tfrac{1}{\sqrt{m}}\sum_{k=1}^m \ket{kk}$. In matrix, $\varsigma^{ab}$ is direct sum of two submatrices, spanned by the local orthonormal bases $\{\ket{kk}, k=1,2, \cdots, m\}$
and $\{\ket{kl}, k\neq l\}$ respectively. However $\varsigma^{ab}$ does not show a "X" form so that an independent discussion is needed.

It should be pointed out that because of the isotropic feature, $\ket{\Psi^+}$ is already in the Schmidt decomposition. The other orthonormal states in the subspace spanned by $\{\ket{kk}, k=1,2, \cdots, m\}$ is written as
\begin{eqnarray}
\ket{\Psi_n^+}=\tfrac{1}{\sqrt{m}}\sum_{k=1}^m e^{i2n\pi \tfrac{k-1}{m}}\ket{kk}, n=1,2, \cdots, (m-1).
\end{eqnarray}
Thus
\begin{eqnarray}
\sqrt{\varsigma^{ab}}&=&\sqrt{x}\ket{\Psi^+}\bra{\Psi^+}+\sqrt{ \frac{1-x}{m^2 -1}}\sum_{n=1}^{m-1} \ket{\Psi_n^+}\bra{\Psi_n^+}\nonumber\\
&& + \sqrt{ \frac{1-x}{m^2 -1}}\sum_{k\neq l} \ket{kl}\bra{kl}.
\end{eqnarray}
Consequently the nearest neighbor $\sigma$ should be the following form
\begin{eqnarray}
\sigma=\sum_{k=1}^m p_k \ket{kk}\bra{kk} +\sum_{k\neq l}p_{kl}\ket{kl}\bra{kl},
\end{eqnarray}
in which $p_k$ and $p_{kl}$ can be decided by Eq. \eqref{bmp}.

By calculation one can obtain
\begin{eqnarray}
&&\max \text{Tr}\left[ \sqrt{\varsigma^{ab} }\sqrt{\sigma} \right]\nonumber \\
&=&\sqrt{m \left[ \tfrac{\sqrt{x}}{m}+\tfrac{m-1}{m}\sqrt{ \tfrac{1-x}{m^2 -1}}\right]^2+(m^2-m) \tfrac{1-x}{m^2 -1} }.
\end{eqnarray}
Thus
\begin{eqnarray}
D^H(\varsigma^{ab})=1- \sqrt{\tfrac{(1-x)(m^2-1)+m}{m(m+1)}+ \tfrac{2}{m}\sqrt{\tfrac{m-1}{m+1}}\sqrt{x(1-x)}},
\end{eqnarray}
which is obviously compatible  with Eq. (17) in Ref. \cite{chang13}
\begin{eqnarray}
D_H(\varsigma^{ab})=1- \left\{\tfrac{(1-x)(m^2-1)+m}{m(m+1)}+ \tfrac{2}{m}\sqrt{\tfrac{m-1}{m+1}}\sqrt{x(1-x)}\right\}.
\end{eqnarray}

In this section we demonstrate the generality and popularity of our conjectures by two examples. We also note that $D^H$ is always less than $D_H$, defined by Eq. (2) in Ref. \cite{chang13}.

\section{Conclusion and  Discussion}

In conclusion, a generalization of the geometric measure of quantum discord is introduced in this article. Our definition can be generalized readily into multipartite case. Moreover since the adopted Hellinger distance and the uninvolved of local measurements, it does not suffered from the critiques raised recently in Refs. \cite{dakic, pz, gheorghiu14}. An important conclusin in this article is that in order to determine the optimal value of Eq. \eqref{def}, it is necessary to find the Schmidt decomposition for the measured state. Then the optimal completely classical state $\sigma$ is a joint  distribution of the corresponding Schmidt basis with the probability decided by Eqs. \eqref{bp} and \eqref{bmp}. In Section III we display the validity of the result by exactly solving several examples. Then two conjectures are presented. Up to our knowledge, it is the first general exact result for the geometric measure of quantum discord.  For multipartite states, the geometric discord can also be evaluated exactly if the state possesses the invariance under permutation or translation, as shown in Section IV. Furthermore it is pointed out in Conjecture \ref{c3} that the optimal $\sigma$ necessarily have  the same invariance.  In Section V we show by two models that our new definition can be used to mark the quantum phase transitions in many-body systems. A discussion of multilevel case is also presented in Section VI. Two examples are worked out exactly, which also are studied in Ref. \cite{chang13}. The fact that our results are compatible with that in Ref. \cite{chang13} unambiguously shows the validity and generality of our conjectures in this article.

Finally we provide a further discussion on our conclusion. As claimed in this article that the optimal $\sigma$ is determined by the Schmidt decomposition of bipartite state, It seems a natural hypothesis  that one could found the generalized Schmidt decomposition for multipartite state based on the "nearest" $\sigma$. Then a geometric understanding of Schmidt decomposition can be constructed by this way, which is inevitably interesting, e.g., in the measure of quantum correlation. Although this approaching is instructive, there are some problems to answer at first. First as for pure bipartite state, the Schmidt decomposition can be used to quantify the quantum entanglement in the state. However it is unclear that this feature is preserved or not when generalized into multipartite. Second as for mixed state, what the meaning of Schmidt decomposition is. We do not know how to understand this point by now. However it is still an interesting way to found the geometric understanding of quantum correlation.

\section*{Acknowledgement}
This work is supported by NSF of China, Grant No. 11005002 (Cui) and 11475004 (Tian), New Century Excellent Talent of M.O.E (NCET-11-0937), and Sponsoring Program of Excellent Younger Teachers in universities in Henan Province of China (2010GGJS-181).

\end{document}